\documentstyle[preprint,prl,aps]{revtex}
\begin{document}
\draft
\title{Effect of Interactions on the Critical Temperature of a Dilute Bose Gas}
\author{M. Bijlsma and H.T.C. Stoof}
\address{Institute of Theoretical Physics, University of
Utrecht, Princetonplein 5, P.O. Box 80.006, 3508 TA Utrecht, The
Netherlands}

\maketitle
\begin{abstract}
Due to the observation of Bose-Einstein condensation in dilute atomic $^{87}$Rb
and $^{23}$Na vapors last year, there is currently a great interest in the
properties of a degenerate Bose gas. One aspect that is not yet understood is
the effect of the interatomic interactions on the critical temperature of the
phase transition. We present here a renormalization group study of the
weakly-interacting Bose gas and predict the critical temperature at which
Bose-Einstein condensation occurs as a function of the (positive) scattering
length $a$ of the interatomic potential.
\end{abstract}

After a long history of trying to master stabilization and cooling, the correct
road towards Bose-Einstein condensation in a dilute gas has now finally been
found. Indeed, last year a Bose condensate was observed in the magnetically
trapped and evaporatively cooled alkali gases $^{87}$Rb and $^{23}$Na
\cite{rub,nat}. Initiated by this breaktrough, extensive experimental and
theoretical studies will surely follow in the near future to fathom the physics
of Bose-Einstein condensation in inhomogeneous gases. Superfluidity
\cite{STRING1}, the Josephson effect and vortex dynamics are just a few
examples of phenomena that are very interesting, but in first instance also
very difficult to experimentally measure or create. Aspects that appear to be
more easily accessible and will be subject of scrutiny first, are e.g.\
collective excitations \cite{EDWBUR} and the critical temperature of the phase
transition. With respect to this latter point, the recent experiments claim to
observe a critical temperature that is shifted upward compared to the ideal gas
value, as the measured degeneracy parameter $n_c\Lambda_{th}^3$ at the
transition is smaller than $\zeta (3/2)\simeq 2.612$. The exact magnitude of
the shift however, is still uncertain. Here $n$ is the density and
$\Lambda_{th}=(2\pi\hbar^2/mk_BT)^{1/2}$ the thermal de Broglie wavelength of
the atoms in the gas.

A shift in the critical temperature can be expected because the gas contains a
finite number of particles. However, this effect is negligible for the large
number of atoms used in the experiments \cite{KETT1}, and would shift the
critical temperature downwards. In addition, a shift can arise because the
atoms are interacting via an (effectively) repulsive potential. Concerning this
latter aspect, it is important to note that existing approaches to the
weakly-interacting Bose gas using the Bogoliubov (or Popov) theory actually do
not predict a shift in the critical temperature \cite{USVAR}. Since these
theories are of mean-field type, one might naively expect that critical
fluctuations will depress the critical temperature, which would again
contradict the experimental data. To settle this issue, and also give a
quantative prediction for the shift of the critical temperature, we report here
on a renormalization group (RG) calculation for the homogeneous Bose gas.

In most applications, the RG method is used to obtain information about
universal properties of the system considered, such as critical exponents.
Nonuniversal properties depend on the (usually unknown) ultraviolet cutoff
$\Lambda$ of the theory. However, in our case we can eliminate this cutoff
dependence because we have sufficient information on the microscopic details of
the dilute Bose system. Therefore, we are also able to determine nonuniversal
properties of the system as we will see shortly.

In principle, we treat here only the homogeneous Bose gas with effectively
repulsive interactions, which implies that the interatomic potential has a
positive scattering length $a$. However, as the number of particles $N\gg 1$,
the critical temperature $T_c$ is much larger than the energy splitting
$\hbar\omega$ between the levels in the harmonic oscillator traps used
experimentally. As the width of the region in which finite-size effects are
important is of $O(T_c(\hbar\omega/k_BT_c))$, one can practically for all
temperatures use a local density approximation to describe the gas in the
center of the trap. As a result, the critical conditions we find here for the
homogeneous gas are essentially also valid for the central density of a trapped
Bose gas, and in particular pertain also to the $^{87}$Rb and $^{23}$Na
experiments.

In the functional formulation of the grand canonical partition function, the
action for the dilute Bose system in momentum and frequency space is written as
\begin{eqnarray}
S[a^*,a]&=&\sum_{{\bf k},n}(-i\hbar\omega_n+\epsilon_{{\bf k}}-\mu)
a^*_{{\bf k},n}a_{{\bf k},n}\nonumber \\
&+&\frac{1}{2}\frac{1}{\hbar\beta V} \sum_{\stackrel{\bf
k,k',q}{n,n',m}} V_{\bf q}a^*_{{\bf k+q},n+m}a^*_{{\bf k'-q},n'-m}
a_{{\bf k'},n'}a_{{\bf k},n}~,
\end{eqnarray}
where $\omega_n=2\pi n/\hbar\beta\equiv 2\pi nk_BT/\hbar$ are the bosonic
Matsubara frequencies, $\epsilon_{\bf k}=\hbar^2{\bf k}^2/2m$, $\mu$ is the
chemical potential, $V$ is the volume of the system, $V_{\bf q}=\int d{\bf
x}~V({\bf x})e^{-i{\bf q\cdot x}}$ is the Fourier transform of the interaction
potential, and $a^*_{{\bf k},n}$ and $a_{{\bf k},n}$ are the Fourier components
of the fields $\psi^*({\bf x},\tau)$ and $\psi({\bf x},\tau)$, creating
respectively annihilating a particle at position ${\bf x}$ and (imaginary) time
$\tau$. The specific form of the potential $V({\bf x})$ or its corresponding
scattering length is in first instance immaterial to the derivation of the RG
equations.

The renormalization of the various terms in the effective action is obtained by
successively integrating out high momentum shells d$\Lambda$ of infinitesimal
width \cite{wilson}. In the operator language this is equivalent to performing
the trace in the grand canonical partition function over the most rapidly
oscillating one-particle states, and then constructing a new effective
Hamiltonian for the less rapidly oscillating states. Technically, this first
step of Wilson's RG transformation boils down to calculating the Feynman
diagrams renormalizing the vertices of interest, where the integration over the
internal momenta is restricted to the afore-mentioned momentum shell. The type
of diagrams contributing to the renormalization when the thickness of the
momentum shell d$\Lambda$ is infinitesimal, are the so-called one-loop diagrams
only. This follows from the fact that each extra loop introduces an extra
factor of d$\Lambda$ \cite{HUANG}. The second step in Wilson's RG
transformation consists of a rescaling of the momenta, frequencies and fields
such that the original cutoff is restored and the coefficient of
$(-i\hbar\omega_n+\epsilon_{\bf k})a^*_{{\bf k},n}a_{{\bf k},n}$ is kept equal
to one. If one neglects the renormalizations from the first step, this
procedure yields the so-called trivial scaling of the vertices and reveals the
relevance of the vertices considered.

In principle, the action in Eq.\ (1) does not possess a real sharp ultraviolet
cutoff $\Lambda$. However, the typical behavior of the Fourier transform of the
two-body interaction potential is such that there is an effective ultraviolet
cutoff around the momentum scale set by the scattering length $a$ of this
potential. Below this value, the Fourier transform is practically momentum
independent and equal to $V_{\bf 0}$. As in the Bose systems considered here we
always have that $\hbar/a\gg\hbar/\Lambda_{th}$, the momentum range in which
the particles in the gas reside falls well below this cutoff. Thus, we
represent the interaction potential by the momentum independent value $V_{\bf
0}$ for $q$ less than a cutoff $\Lambda$ of $O(1/a)$ and zero for $q>\Lambda$.

Modelling the potential as described, implies that the nonuniversal properties
we find from an RG calculation will in first instance be sensitive to the
specific value of the cutoff $\Lambda$ taken in the calculations. However, at
this point our knowledge about the microscopic details of the Bose gas comes in
to resolve this problem because we know that the two-body interaction potential
$V_{\bf q}$ has to renormalize to the two-body $T-$matrix $T^{2B}(({\bf
k-k'})/2+{\bf q},({\bf k-k'})/2;\hbar^2({\bf k-k'})^2/m)$ when we include all
possible two-body scattering processes in the vacuum. The two-body $T-$matrix
has roughly the same momentum dependence as $V_{\bf q}$, and is in particular
constant and equal to $4\pi a\hbar^2/m$ in the range of thermal momenta and
energies. Thus, given an ultraviolet cutoff $\Lambda$, we can fix the RG
equations by demanding that for the two-body problem $V_{\bf 0}$ indeed
correctly renormalizes to $4\pi a\hbar^2/m$. Since this value is, due to the
inequality $a/\Lambda_{th}\ll 1$, already attained before we enter into the
thermal regime as we integrate out more and more momentum shells, this indeed
leads to a correct description of the properties of the Bose gas which is
independent of the ultraviolet cutoff $\Lambda$. Having eliminated the cutoff
dependence in this manner, we are then in a position to determine also the
nonuniversal properties of the dilute Bose gas such as most notably the
critical temperature. Furthermore, we can perform the calculation for any
(positive) value of the scattering length, thus being able to describe any
atomic species with effectively repulsive $s-$wave scattering. The results we
find are therefore relevant to the experiments using $^{87}$Rb, $^{23}$Na, but
for instance also for future experiments with $^1$H and other atoms having a
positive scattering length.

When deriving the equations governing the renormalization of the various terms
in the action, one has to proceed differently, depending on the sign of the
chemical potential. For negative chemical potential the derivation of the RG
equations is most straightforward as one can start directly from Eq.\ (1) to
obtain the Feynman diagrams. For positive chemical potential the derivation is
much more involved since in this case the space and time independent part of
the effective action, i.e.\ $-\mu\mid a_{{\bf 0},0}\mid^2+V_{\bf 0}\mid a_{{\bf
0},0}\mid^4/2\hbar\beta V$, has a Mexican hat shape and one first has to expand
the action around the correct extremum by performing the shift $a_{{\bf
0},0}\rightarrow a_{{\bf 0},0}+\sqrt{n_{\bf 0}\hbar\beta V}$ introducing the
condensate density $n_{\bf 0}$. Only after that can one find the Feynman
diagrams contributing to the renormalization.

We will first focus on the situation of negative chemical potential, and
restrict ourselves to operators which are relevant or marginal at the critical
temperature. The chemical potential $\mu$ and the two-body interaction term
$V_{\bf 0}$ are both relevant there. The vertex $U_{\bf 0}$ from the three-body
interaction term, schematically written as $U_{\bf 0}\mid\psi({\bf
x},\tau)\mid^6/6$ and in principle also present in the Hamiltonian of the
system, is marginal, and would thus have to be included in the set of RG
equations describing the change or `flow' of the vertices during the process of
integrating out momentum shells. However, since we are studying the {\em
dilute} Bose gas it it expected that three-body interactions will in general
have only an extremely small influence on the properties to be calculated. Only
near the critical temperature can there be some effect as at that point all
fluctuations become important. This we have checked explicitly by numerically
solving the coupled set of RG equations including and excluding the three-body
interaction. Indeed, the above argument is corroborated by our findings, which
show that the influence of $U_{\bf 0}$ on e.g.\ the density, even in the
critical region, is typically smaller than 1\% in the regime where $na^3\ll 1$.
In the following we will, therefore, neglect the three-body interaction and
focus on the renormalization of $\mu$ and $V_{\bf 0}$ only. For negative
chemical potential, the RG equations are then given by
\begin{mathletters}
\begin{eqnarray}
\frac{d\mu}{dl}=2\mu-\frac{\Lambda^3}{\pi^2}V_{\bf 0}N(\epsilon_{\Lambda} -\mu)
\end{eqnarray}
\begin{eqnarray}
\frac{dV_{\bf 0}}{dl}&=&-V_{\bf 0}-\frac{\Lambda^3}{2\pi^2}V_{\bf 0}^2
\frac{1+2N(\epsilon_{\Lambda}-\mu)}{2(\epsilon_{\Lambda}-\mu)}\nonumber \\
&&-\frac{2\Lambda^3}{\pi^2}V_{\bf 0}^2\beta
N(\epsilon_{\Lambda}-\mu)[N(\epsilon_{\Lambda}-\mu)+1]~.
\end{eqnarray}
\end{mathletters}
At a particular value of $l$, the region in momentum space that has been
integrated out is the shell running from $\mid{\bf k}\mid=\Lambda e^{-l}$ to
$\mid {\bf k}\mid=\Lambda$. Here,
$N(\epsilon_{\Lambda}-\mu)=1/(e^{\beta(\epsilon_{\Lambda} -\mu)}-1)$ is the
Bose-Einstein distribution function containing
$\epsilon_{\Lambda}=\hbar^2\Lambda^2/2m$ in its argument since after each step
we have rescaled the momenta to retain the cutoff at $\Lambda$. Furthermore, as
a result of this rescaling we have that the temperature scales as
$T(l)=Te^{2l}$. The nontrivial term in Eq.\ (2a) originates from the sum of the
Hartree and Fock diagrams, whereas the nontrivial terms in Eq.\ (2b) stem from
the ladderdiagram and the bubblediagrams respectively.

Next we consider the case of positive chemical potential. Introducing the
condensate density through the shift mentioned above, we generate, next to a
linear term proportional to $(a_{{\bf 0},0}+a^*_{{\bf 0},0})$, normal and
anomalous selfenergies $\sum_{{\bf k},n}(\hbar\Sigma_{11}a^*_{{\bf k},n}a_{{\bf
k},n}+\hbar\Sigma_{12}(a^*_{{\bf k},n}a^*_{-{\bf k},-n}+a_{{\bf k},n}a_{-{\bf
k},-n})/2)$ and the term $\sum_{{\bf k},n}\sum_{{\bf q},m}\Gamma_3(a^*_{{\bf
q},m}a^*_{{\bf k-q},n-m}a_{{\bf k},n}+a^*_{{\bf k+q},n+m}a_{{\bf q},m}a_{{\bf
k},n})/\sqrt{\hbar\beta V}$ into the action \cite{USVAR}. In first instance we
have, due to the neglect of the three-body interactions, that
$\hbar\Sigma_{11}=2n_{\bf 0}V_{\bf 0}$, $\hbar\Sigma_{12}=n_{\bf 0}V_{\bf 0}$,
and $\Gamma_3=\sqrt{n_{\bf 0}}V_{\bf 0}$. Moreover, the magnitude of the
condensate is found by eliminating the linear term in the action, which gives
$n_{\bf 0}=\mu/V_{\bf 0}$.

Performing a Bogoliubov transformation to diagonalize the quadratic part of the
action facilitates the calculation of the one-loop diagrams contributing to the
renormalization of the vertices of interest. One of these corresponds to the
linear term in the action. After every step of the RG transformation we have to
eliminate this term by performing a shift in $a_{{\bf 0},0}$ in order to remain
in the minimum of the action. Clearly this gives the equation describing the
change of the condensate density. Proceeding with the selfenergies, we find
that they obey the Hugenholtz-Pines relation
$\mu(l)=\hbar\Sigma_{11}(l)-\hbar\Sigma_{12}(l)$ implicate from the $U(1)$
symmetry of the problem \cite{HUPI}. As a result, we only have to consider the
diagrams contributing to $\hbar\Sigma_{12}$ to obtain the renormalization of
the quadratic part of the action. Finally, due to the neglect of the three-body
interactions, we have again from $U(1)$ symmetry that $\Gamma_3(l)=\sqrt{n_{\bf
0}(l)}V_{\bf 0}(l)$ and $V_{\bf 0}(l)=\hbar\Sigma_{12}(l)/n_{\bf 0}(l)$. These
relations allow the action to be recast into the explicitly $U(1)$ symmetric
form of Eq.\ (1) at any point during renormalization. Consequently, from the RG
equations for $\hbar\Sigma_{12}$ and $n_{\bf 0}$, we can deduce the equations
describing the flow of the chemical potential $\mu$ and the two-body
interaction $V_{\bf 0}$. This leads to
\begin{mathletters}
\begin{eqnarray}
\frac{d\mu}{dl}&=&2\mu-\frac{\Lambda^3}{2\pi^2}V_{\bf
0}\left[\frac{2\epsilon_{\Lambda}^3+6\mu\epsilon_{\Lambda}^2+
\mu^3}{2\hbar^3\omega_{\Lambda}^3}(2N(\hbar\omega_{\Lambda})+1)-1\right.
\nonumber \\
&&+\left.\frac{\mu(2\epsilon_{\Lambda}+\mu)^2}{\hbar^2\omega_{\Lambda}^2} \beta
N(\hbar\omega_{\Lambda})[N(\hbar\omega_{\Lambda})+1]\right]
\end{eqnarray}
\begin{eqnarray}
\frac{dV_{\bf 0}}{dl}&=&-V_{\bf 0}-\frac{\Lambda^3}{2\pi^2}V_{\bf 0}^2
\left[\frac{(\epsilon_{\Lambda}-\mu)^2}{2\hbar^3\omega_{\Lambda}^3}
(2N(\hbar\omega_{\Lambda})+1)\right.\nonumber \\
&&+\left.\frac{(2\epsilon_{\Lambda}+\mu)^2}{\hbar^2\omega_{\Lambda}^2} \beta
N(\hbar\omega_{\Lambda})[N(\hbar\omega_{\Lambda})+1]\right]~,
\end{eqnarray}
\end{mathletters}
where the Bose-Einstein distribution function now contains the Bogoliubov
dispersion $\hbar\omega_{\Lambda}= \sqrt{\epsilon_{\Lambda}^2
+2\mu\epsilon_{\Lambda}}$. Eqs.\ (2) and (3) constitute the flow equations in
the case of negative and positive chemical potential respectively, and
completely descibe the system in the dilute gas approximation. Note that both
sets coincide when $\mu$ is taken equal to zero, implying that the flow is
continuously differentiable, also at $\mu=0$.

Finally, we have to fix the RG equations as explained above. From Eq.\ (2b) we
recognize that in a vacuum, i.e. $N(\epsilon_\Lambda-\mu)=0$, the
renormalization of the interaction between two particles is governed by
\begin{eqnarray}
\frac{dV_{\bf 0}}{dl}=-V_{\bf 0}-\frac{\Lambda^3}{2\pi^2}V_{\bf
0}^2\frac{1}{2(\epsilon_{\Lambda}-\mu)}~.
\end{eqnarray}
As expected, this is just the differential form of the Lippmann-Schwinger
equation for the two-body $T-$matrix at energy $2\mu$, $T^{2B}({\bf 0},{\bf
0};2\mu)$. As the two-body $T-$matrix is energy independent for low energies,
we can neglect $\mu$ and use $T^{2B}({\bf 0},{\bf 0};2\mu)=4\pi a\hbar^2/m$.
The requirement is now that, given an ultraviolet cutoff $\Lambda$, $V_{\bf 0}$
flows to the value $4\pi a\hbar^2/m$. This can be ascertained by choosing the
right initial condition for $V_{\bf 0}$. More precisely we find from Eq.\ (4)
that $V_{\bf 0}(l=0)=4\pi a\hbar^2/(m(1-2a\Lambda/\pi))$ leads to the correct
result. Clearly, we can describe different atomic species by only changing the
value of the scattering length $a$ used in this equation. Note again that this
boundary value also assures the correct renormalization of the full RG
equations (2)-(3), because the other terms in the right-hand side of these
equations are insensitive to the cutoff. Put differently, these terms only
start to contribute to the flow in the range of thermal momenta, but at that
point $V_{\bf 0}$ has already correctly renormalized to $4\pi a\hbar^2/m$.

Having found the correct boundary conditions, $\mu(l=0)$ being the bare
chemical potential in Eq.\ (1), we can now numerically integrate the RG
equations. In Fig.\ 1 we depict the trajectories resulting from these
calculations. The temperature is fixed and the bare chemical potential is
raised going from a lower curve to a higher curve. We can determine the
corresponding density by noting that $n=n_{\bf 0}+\sum_{{\bf k},n}\langle
a^*_{{\bf k},n}a_{{\bf k},n}\rangle$, which we can also cast into a
differential equation yielding the building up of the density as we integrate
out more and more momentum shells. The trajectory flowing into the fixed point
$(\mu^*,V_{\bf 0}^*)$ corresponds after removal of the trivial scaling to a
chemical potential renormalizing exactly to zero, and therefore to a condensate
density renormalizing exactly to zero. For values higher than this (positive)
bare critical chemical potential we find $n_{\bf 0}>0$, and we are in the
condensed phase.

Thus, the critical density of the dilute Bose system is determined by the set
Eq.\ (3). Moreover, we can find the usual critical exponent $\nu$ pertaining to
the divergence of the correlation length as we approach the critical
temperature, i.e.\ $\xi$ behaves as $\xi_0\mid (T-T_c)/T_c\mid^{-\nu}$, by
linearizing the flow equations around the fixed point and calculating the
largest eigenvalue. This gives $\nu=0.685$, which is to be compared with the
value $\nu=0.67$ found from an $\epsilon -$expansion of the $O(2)$ model
\cite{ZINNJ} and also measured in $^4$He experiments. The agreement is
surprisingly good, indicating that we are indeed accurately describing the Bose
gas with the derived RG equations, also in the critical region.

We now turn to the ultimate goal of this study, namely the effect of
interactions on the critical temperature of Bose-Einstein condensation. At
fixed temperatures we vary the (bare) chemical potential to find the
trajectories flowing into the fixed point. As mentioned above, this yields the
critical densities for these specific temperatures and gives us the $n_c-T$
relation. We repeat this for different values of the scattering length to
obtain the dependence of the critical temperature on the strength of the
interaction. All curves fall on one single curve if we plot the degeneracy
parameter $n_c\Lambda_{th}^3$ at which Bose-Einstein condensation occurs versus
$a/\Lambda_{th}$. This curve is depicted in Fig.\ 2, where we normalized
$n_c\Lambda_{th}^3$ to the ideal gas value $\zeta (3/2)$. As seen from Fig.\ 2
we always have that $n_c\Lambda_{th}^3<\zeta (3/2)$, which leads to the
conclusion that the critical temperature of the weakly-interacting Bose gas is
{\em raised} with respect to the ideal gas value, in qualitative agreement with
the recent experiments. The main reason for the lower value of
$n_c\Lambda^3_{th}$ is that at the critical point, the effective chemical
potential renormalizes from a positive initial value exactly to zero. Due to
the Bogoliubov dispersion, this behavior depresses the occupation of the
non-zero momentum states relative to the ideal gas case.

In summary, we have derived renormalization group equations for the dilute Bose
gas. Using our knowledge of the two-body scattering problem, we have
effectively eliminated the ultraviolet cutoff from our theory, and obtained
also information on nonuniversal properties. As such, we have determined the
effect of repulsive interactions on the critical temperature of Bose-Einstein
condensation, and found that it is raised with respect to the ideal gas value.
Preliminary Quantum Monte Carlo calculations seem to confirm this result
\cite{CEPERL}. From our calculations we predict that for the $^{87}$Rb and
$^{23}$Na experiments the critical temperature can be raised with as much as 10
\%. This appears to be a very promising result because one might expect that an
effect of this magnitude can very well be measured in future, more accurate,
experiments.

We acknowledge helpful discussions with Eric Cornell, Wolfgang Ketterle and
Steve Girvin.

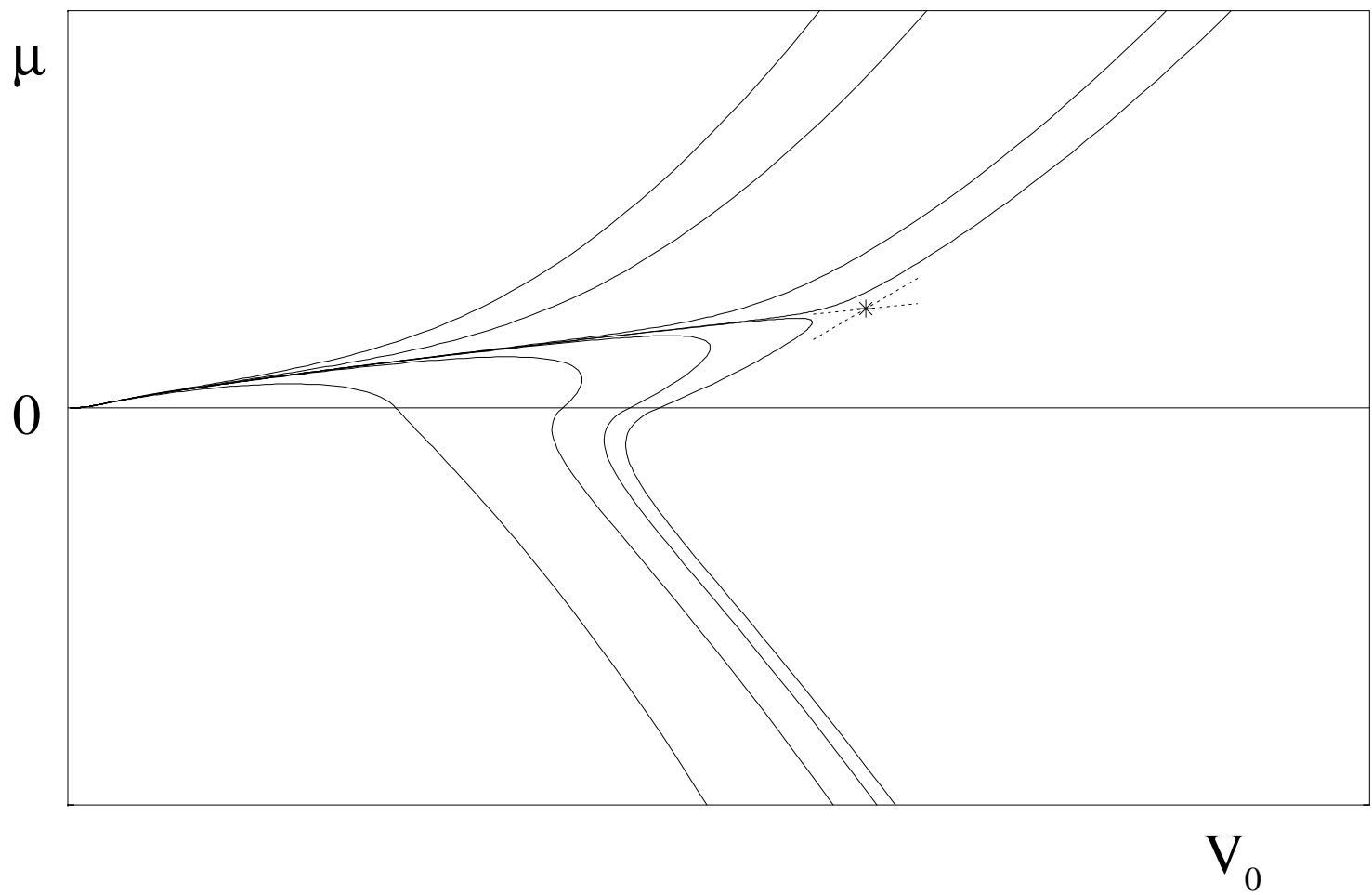
\begin{figure}[h]
\caption{Flowdiagram resulting from the RG equations. The fixed point is
indicated with an asterix and its eigenvectors by the dotted lines.}
\end{figure}
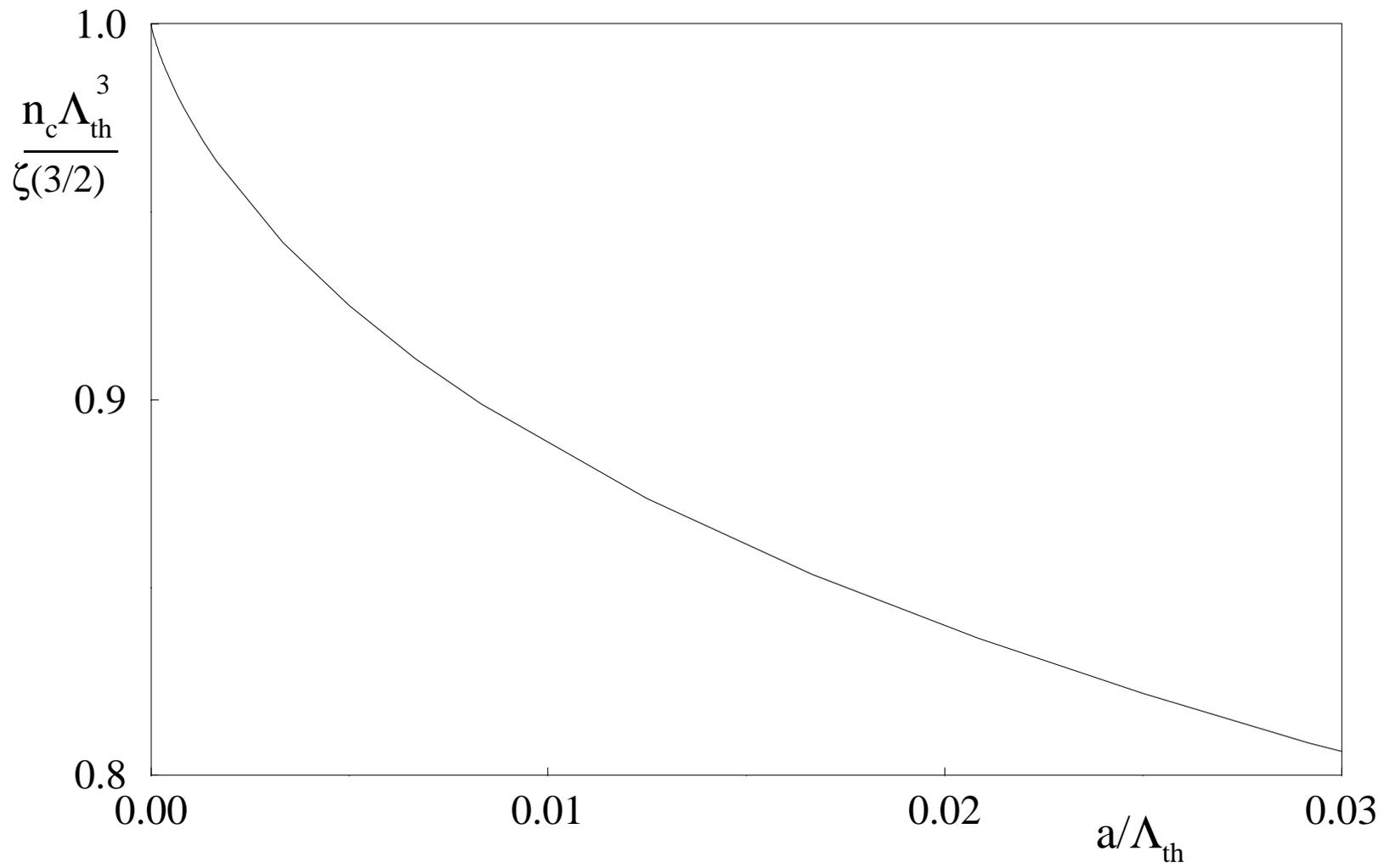
\begin{figure}[h]
\caption{The normalized critical degeneracy parameter versus $a/\Lambda_{th}$.}
\end{figure}
\end{document}